\documentclass[pdflatex,sn-mathphys-ay]{sn-jnl}% Math and Physical Sciences Author Year Reference Style
\usepackage{graphicx}%
\usepackage{multirow}%
\usepackage{amsmath,amssymb,amsfonts}%
\usepackage{amsthm}%
\usepackage{mathrsfs}%
\usepackage[title]{appendix}%
\usepackage{xcolor}%
\usepackage{textcomp}%
\usepackage{manyfoot}%
\usepackage{booktabs}%
\usepackage{algorithm}%
\usepackage{algorithmicx}%
\usepackage{algpseudocode}%
\usepackage{listings}%
\usepackage{comment}
\usepackage{natbib}  % optional, for author-year citation styles

\usepackage{tabularx}   % allows tabularx environment
\usepackage{float}        % for [H] specifier (forces table placement "Here")
\usepackage{makecell}     % for \makecell to allow line breaks in table cells
\usepackage{array}        % improves column formatting (recommended)
\newcolumntype{P}[1]{>{\rule{0pt}{1.0em}}p{#1}}

\theoremstyle{thmstyleone}%
\theoremstyle{thmstyletwo}%

\theoremstyle{thmstylethree}%

\begin{document}

\title[Article Title]{A Benchmark Study of Classical and Dual Polynomial Regression (DPR)-Based Probability Density Estimation Techniques}


\author*[1,3]{\fnm{Shantanu} \sur{Sarkar}}\email{shantanu75@gmail.com}

\author[2,3]{\fnm{Mousumi} \sur{Sinha}}\email{mousumi.sinha@uth.tmc.edu}

\author[3]{\fnm{Dexter} \sur{Cahoy}}\email{cahoyd@uhd.edu}

\affil*[1]{\orgdiv{Dept. of ECE}, \orgname{University of Houston}, \orgaddress{\city{Houston}, \state{TX}, \country{USA}}}

\affil[2]{\orgname{University of Texas Health Science Center}, \orgaddress{\city{Houston}, \state{TX}, \country{USA}}}
\affil[3]{\orgdiv{College of Sciences and Tech.}, \orgname{Univ. of Houston–Downtown}, \orgaddress{\state{TX}, \country{USA}}}

% Computational Statistics
% Please provide an abstract of 150 to 250 words. The abstract should not contain any undefined abbreviations or unspecified references.
% 247 Word
\abstract{The probability density function (PDF) plays a central role in statistical and machine learning modeling. Real-world data often deviates from Gaussian assumptions, exhibiting skewness and exponential decay. To evaluate how well different density estimation methods capture such irregularities, we generated six unimodal datasets from diverse distributions that reflect real-world anomalies. These were compared using parametric methods (Pearson Type I and Normal distribution) as well as non-parametric approaches, including histograms, kernel density estimation (KDE), and our proposed method. To accelerate computation, we implemented GPU-based versions of KDE (tKDE) and histogram estimation (tHDE) in TensorFlow, both of which outperform Python SciPy’s KDE. Prior work demonstrated the use of piecewise modeling for density estimation, such as local polynomial regression; however, these methods are computationally intensive. Based on the concept of piecewise modeling, we developed a computationally efficient model, the Dual Polynomial Regression (DPR) method, which leverages tKDE or tHDE for training. DPR employs the piecewise strategy to split the PDF at its mode and fit polynomial regressions to the left and right halves independently, enabling better capture of the asymmetric shape of the unimodal distribution. We used the Mean Squared Error (MSE), Jensen-Shannon Divergence (JSD), and Pearson’s correlation coefficient, with reference to the baseline PDF, to validate accuracy. We verified normalization using Area Under the Curve (AUC) and computational overhead via execution time. Validation on real-world systolic and diastolic data from 300,000 unique patients shows that the DPR of order 4, trained with tKDE, offers the best balance between accuracy and computational overhead.}

\keywords{PDF, KDE, Histogram, Piecewise, Regression, Computational efficiency}
\pacs[MSC Classification]{62G05,62G08}


\maketitle

\section{Introduction}\label{sec1}
Estimating probability density functions (PDFs) from a dataset is the foundational task in statistical and machine learning modeling. In practical applications, for the ease of computation, we usually assume that data follow a Gaussian distribution; however, real-world data often deviates from this assumption, exhibiting skewness, exponential decay, heavy tails, multimodal characteristics, and contamination from outliers. These deviations challenge standard parametric approaches and highlight the need for more flexible and adaptive density estimation techniques \citep{Ruzgas2025}.
\\% -- Next Paragraph -----------------------------------------
A variety of PDF estimation methods are available, but non-parametric techniques such as histograms and Kernel Density Estimation (KDE) are widely used due to their simplicity and minimal assumptions. KDE is generally preferred compared to the histogram approach for density estimation because it produces a smooth, continuous estimate that is not sensitive to arbitrary bin edges and better captures the underlying structure of the data \citep{Silverman1998, Wglarczyk2018}. The KDE is resource-intensive, which makes it a prohibitively impractical approach for large datasets \citep{Raykar2010}. Parametric approaches for PDF estimation using the Normal distribution or Pearson distribution family provide interpretable models but rely on stronger assumptions about data shape, which may not always hold \citep{R2024}.
\\% -- Next Paragraph -----------------------------------------
\noindent In recent years, piecewise modeling has been advocated for density estimation \citep{Grosdos2023, Zhao2025}. The R package \texttt{`lpdensity'} is an example of the Local polynomial regression, which fits local polynomials at multiple points to capture smooth features of the underlying density \citep{Cattaneo2022}. Building on the insights of piecewise modeling to capture the geometry of the unimodal density function, we propose a computationally efficient Dual Polynomial Regression (DPR) approach that segments the estimated density at its peak and fits separate polynomials to the left and right halves of a unimodal distribution. DPR reduces the evaluation cost to O($M\cdot p$) across M grid points (after $p^{th}$ order polynomial fit on N samples), in contrast, local polynomial density estimation requires O($p^2N$) operations per evaluation point due to weighted least squares, resulting in a total cost of O($M\cdot p^2N$) when evaluated at M grid points \citep{Cattaneo2022}. Prior work on density estimation has often been validated only on limited datasets, using either theoretical analysis or synthetic data, and lacked detailed comparisons against traditional methods. To address this gap, we benchmark DPR against classical approaches, including histograms, kernel density estimation, and parametric fits (Normal and Pearson Type I distributions). For this comparison, we excluded local polynomial density estimation functions, as our focus is on lightweight and computationally efficient alternatives.
\\% -- Next Paragraph -----------------------------------------
\noindent In this work, we benchmark a suite of PDF estimation methods on synthetically generated unimodal datasets such as a Gaussian distribution with and without skewness, asymmetric distributions like Asymmetric M-Wright I (AMW-I), Asymmetric M-Wright II (AMW-II) \citep{Mainardi2010, Cahoy2015}, and an Asymmetric Laplace distribution \citep{Kozubowski2000, Kotz2001} to mimic common statistical properties observed in real-world scenarios, including skewness and asymmetricity within a unimodal distribution. Synthetic datasets were generated from the baseline PDFs by applying inverse transform sampling.
\\% -- Next Paragraph -----------------------------------------
\noindent To benchmark different methods, we performed a comparative analysis of estimation accuracy and computational overhead. The Jensen-Shannon Divergence (JSD) is ideal for comparing two PDFs from the same dataset because it is symmetric, bounded, and measures the similarity between distributions without assuming one as ground truth \citep{Nielsen2019}. Additionally, the Mean Squared Error (MSE) measures the average of the squared point-wise differences, placing a greater penalty on larger errors. Also, the Pearson correlation coefficient measures the linear relationship between two distributions \citep{Dodge2008}. Hence, estimation accuracy is measured via MSE, JSD, and the Pearson correlation coefficient relative to the baseline PDF. To assess computational overhead, we analyzed the time and space complexity of each function theoretically $(O(\cdot))$ and validated the results by recording actual execution times. Additionally, we verified that the PDF integrates to one, confirming normalization.
\\% -- Next Paragraph -----------------------------------------
\noindent Furthermore, we tested our DPR function using a real-world dataset of systolic and diastolic vital signs from 300,000 unique patients over a three-year period, retrieved from the Epic Electronic Health Record System \citep{Epic}. We validated the probability estimation for both the entire dataset and for a small sample size, using a sliding window with 1000 samples. Our results highlight the trade-offs between simplicity, flexibility, and computational cost in probability density estimation. In particular, the proposed Dual Polynomial Regression (DPR) method performed especially well on asymmetric distributions with skewness and long, exponentially decaying tails, making it a promising alternative to classical approaches when data deviates from Gaussian assumptions.
%@@@@@@@@@@@@@@@@@@@@@@@@@@@@@@@@@@@@@@@@@@@@@@@@@@@@@@@@@@@@@@
\section{Methods}\label{sec2}
%@@@@@@@@@@@@@@@@@@@@@@@@@@@@@@@@@@@@@@@@@@@@@@@@@@@@@@@@@@@@@@
\subsection{Dataset}\label{sec2_1}
%**************************************************************
\begin{figure}[t]
\centering
\includegraphics[width=\textwidth]{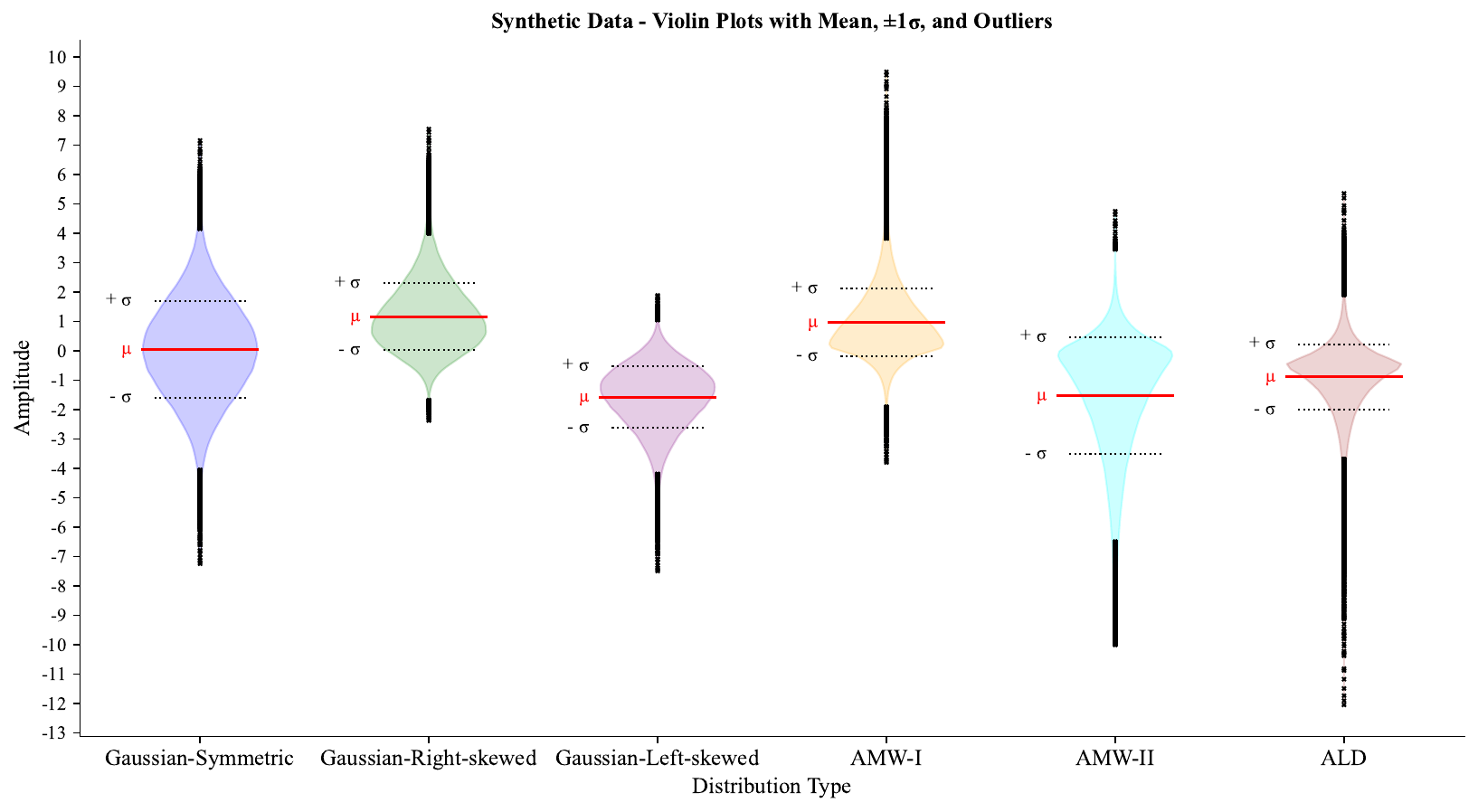}
\caption{Violin plots showing distributions of six synthetic datasets, each represented by different color codes: (i) Symmetric Gaussian (Blue), (ii) Gaussian-Right-skewed (Green), (iii) Gaussian-Left-skewed (Purple), (iv) Asymmetric M-Wright Type-I (AMW-I) (Orange), (v) Asymmetric M-Wright Type-II (AMW-II) (Cyan), and (vi) Asymmetric Laplace Distribution (ALD) (Brown). On each violin plot, the red line indicates the mean, and the dotted black lines mark the $\pm$ standard deviation.}
\label{fig1}
\end{figure}
%**************************************************************
\subsubsection{Synthetic Data}\label{sec2_1_1}
%==============================================================
To evaluate the robustness and flexibility of various probability density estimation techniques - how well they can capture statistical characteristics commonly encountered in real-world scenarios, we generated synthetic datasets for six unimodal distributions: (1) Symmetric Gaussian, (2) Right-skewed Gaussian, (3) Left-skewed Gaussian, (4) Asymmetric M-Wright 1 (AMW-I), (5) Asymmetric M-Wright 1I (AMW-II), and (6) Asymmetric Laplace distribution (ALD). 
\\% -- Next Paragraph -----------------------------------------
For each distribution type, we generated the baseline probability density function (PDF) over a fixed domain spanning [$- 10,\ +10$], and used inverse transform sampling to synthesize 200,000 data points per distribution. For the ALD, we employed piecewise inverse transform sampling by splitting the distribution at a probability threshold ($p$) and applying separate inverse CDF formulas to the left and right tails. Fig.~\ref{fig1} shows the violin plots of the six synthetic datasets generated for benchmarking, providing a graphical overview of their distributions, which are discussed in detail below.
\\% -- Next Paragraph -----------------------------------------
\\% -- Add Gap --------------------------------------------
\textbf{Symmetric Gaussian Distribution: }A standard Gaussian distribution is generated with a random mean value between $\pm0.5$, a variance using a random value between 2 and 3, and a unity shape factor ($\alpha=1$), serving as a baseline for symmetric bell-shaped data.
\\% -- Next Paragraph -----------------------------------------
\\% -- Add Gap --------------------------------------------
\textbf{Right- and Left-Skewed Gaussian Distribution: }Similar to the standard Gaussian distribution, we used a random mean and random variance and introduced a shape factor ($\alpha$) using a random value between 2 and 3. For the right-skewed distribution, we made the shape factor negative, whereas for the left-skewed distribution, we used the shape factor as positive. The Gaussian distribution, both symmetric and asymmetric(skewed), can be described by Equations (\ref{eq:equ1})-(\ref{eq:equ3}) \citep{Papoulis2002,Azzalini2013}.
%-------------------------------------------------------------
\begin{equation}
f(x; \mu, \sigma, \alpha) = \frac{2}{\sigma} \, \varphi(z) \, \Phi(\alpha z)
\label{eq:equ1}
\end{equation}
%-------------------------------------------------------------
\begin{equation}
\varphi(z) = \frac{1}{\sqrt{2\pi}} e^{-z^2/2}
\label{eq:equ2}
\end{equation}
%-------------------------------------------------------------
\begin{equation}
\Phi(\alpha z) = \frac{1}{\sqrt{2\pi}} \int_{-\infty}^{\alpha z} e^{-t^2/2} \, dt
\label{eq:equ3}
\end{equation}
%-------------------------------------------------------------
Here:
\begin{flushleft}
\( z = \frac{x - \mu}{\sigma} \): standard normal variable.\\[0.5em]
\( \varphi(z) \): standard normal PDF.\\[0.5em]
\( \Phi(z) \): standard normal CDF.\\[0.5em]
\( \mu \): mean.\\[0.5em]
\( \sigma \): standard deviation.\\[0.5em]
\( \alpha \): shape factor,  \(
\alpha = 0 \text{ (symmetric)}, \quad
\alpha > 0 \text{ (right-skewed)}, \quad
\alpha < 0 \text{ (left-skewed).}
\)
\end{flushleft}
%=============================================================
\textbf{Asymmetric M-Wright Type I (AMW-I) and Type II (AMW-II):}
Anomalous diffusion, represented by time-fractional diffusion equations, is better captured by the M-Wright function, also known as the Mainardi function \citep{Mainardi2010}, as mathematically represented by Equation (\ref{eq:equ4}). Using the M-Wright function, two asymmetric unimodal distribution functions are defined: Asymmetric-M-Wright Type-I (AMW-I) and Asymmetric M-Wright Type II (AMW-II ) \citep{Cahoy2015}, as described by Equations (\ref{eq:equ5}) and (\ref{eq:equ7}), respectively. We employed the M-Wright function with the Gamma function reflection formula \citep{Cahoy2015}, as shown in Equation (\ref{eq:equ8}), which helped avoid singularities in the original denominator and improved numerical stability by avoiding the Gamma function of negative values.
\\% -- Next Paragraph -----------------------------------------
For both AMW-I and AMW-II, while generating the distribution function, we randomly selected the shape parameter ($\nu$) from the range 0.15 to 0.35. For both AMW-I and AMW-II, we used positive random Skewness Parameter ($\lambda$) and ($\alpha$), respectively, where $\lambda \geq1,\ \alpha  \ \leq2$.
%-------------------------------------------------------------
\begin{equation}
M_\nu(x) = \sum_{j=0}^{\infty} \frac{(-x)^j}{j! \, \Gamma(-\nu j + 1 - \nu)}\\[0.5em]
\label{eq:equ4}
\end{equation}
%-------------------------------------------------------------
\begin{equation}
f_\nu(x; \lambda) = M_\nu(|x|) \cdot \mathfrak{M}_\nu(\lambda x), \quad 0 \leq \nu < 1\\[0.5em]
\label{eq:equ5}
\end{equation}
%-------------------------------------------------------------
\begin{equation}
\mathfrak{M}_\nu(y) = \frac{1}{2} \left[ 1 + \operatorname{sgn}(y) \left( 1 - \sum_{j=0}^{\infty} \frac{(-|y|)^j}{j! \, \Gamma(1 - \nu j)} \right) \right]\\[0.5em]
\label{eq:equ6}
\end{equation}
%-------------------------------------------------------------
\begin{equation}
f_\nu(x; \alpha) = \frac{\alpha}{1 + \alpha^2}
\begin{cases}
M_\nu(\alpha x), & \text{if } x \geq 0 \\
M_\nu\left(-\frac{x}{\alpha}\right), & \text{if } x < 0\\[0.5em]
\end{cases}
\label{eq:equ7}
\end{equation}
%-------------------------------------------------------------
\begin{equation}
M_\nu^R(x) = \frac{1}{\pi} \sum_{j=0}^{\infty} \frac{(-x)^j \, \Gamma(\nu j + \nu) \, \sin\left( \pi(\nu j + \nu) \right)}{j!}\\[0.5em]
\label{eq:equ8}
\end{equation}
%-------------------------------------------------------------
Here, 
\begin{flushleft}
\(\nu \in (0, \tfrac{1}{2})\): Shape parameter for unimodal distribution (controls tail spread).\\[0.3em]
%-----------------------------------------------------------
\(x \in \mathbb{R}\): Real-valued input (evaluation point).\\[0.3em]
%-----------------------------------------------------------
\(j \in \mathbb{N}_0\): Summation index (non-negative integers).\\[0.3em]
%-----------------------------------------------------------
\(\Gamma(\cdot)\): Gamma function \(\left( \Gamma(n) = (n-1)! \text{ for } n \in \mathbb{N} \right)\).\\[0.3em]
%-----------------------------------------------------------
\(M_\nu(|x|)\): M-Wright function of \(|x|\).\\[0.3em]
%-----------------------------------------------------------
\(\mathfrak{M}_\nu(\lambda x)\): CDF associated with the symmetric M-Wright PDF.\\[0.3em]
%-----------------------------------------------------------
\(\lambda \in \mathbb{R}\): Skewness parameter — governs asymmetry and skewness in AMW-I.\\[0.3em]
\(\alpha \in \mathbb{R}^+\): Skewness parameter — governs asymmetry and skewness in AMW-II.
\end{flushleft}
%-----------------------------------------------------------
\textbf{Asymmetric Laplace Distribution (ALD): } The Asymmetric Laplace Distribution (ALD) is a continuous distribution that extends the classic Laplace distribution by allowing different decay rates on either side of its central location. As in a real-life scenario, a good number of datasets, such as survival analysis in healthcare \citep{Sheng2025}, financial data \citep{Jing2022,Punzo2025}, etc., are precisely modeled utilizing ALD, so we also considered analyzing ALD separately, even though the Asymmetric M-Wright distribution with adjusted parameters can behave as ALD \citep{Cahoy2015}. ALD function is mathematically explained by Equation (\ref{eq:equ9}) \citep{Kotz2001,Punzo2025}. 
\\% -- Next Paragraph -----------------------------------------
In our study, we generated the ALD by randomly picking location parameter ($m$), scale factor ($\lambda$), and asymmetry factor ($\kappa$) from the range: $m \in [ -0.5,0.5 ],\ \lambda \in [ 1,2 ],\ \text{and}\ \kappa \in [ 1.2,1.8 ]$, respectively.
%-----------------------------------------------------------
\begin{equation}
f(x \mid m, \sigma, \kappa) = \frac{\lambda \kappa}{(1 + \kappa)^2}
\begin{cases}
\exp\left(-\frac{\lambda}{\kappa} |x - m|\right), & \text{if } x < m \\
\exp\left(-\frac{\kappa}{\lambda} |x - m|\right), & \text{if } x \geq m
\end{cases}
\label{eq:equ9}
\end{equation}
%-----------------------------------------------------------
Here, 
\begin{flushleft}
$m \in \mathbb{R}$: Location parameter.\\[0.5em]
%-----------------------------------------------------------
$\lambda \in \mathbb{R}^+$: Scale factor.\\[0.5em]
%-----------------------------------------------------------
$\kappa \in \mathbb{R}^+$: Asymmetry factor ($=1$: symmetric; 
$<1$: right-skewed; 
$>1$: left-skewed).
%-----------------------------------------------------------
\end{flushleft}
\textbf{Inverse Transform Sampling: }Inverse Transform Sampling is a method for generating random samples from a PDF, by normalizing the PDF utilizing step size and then mapping uniformly distributed random numbers through the inverse computed CDF to obtain samples that follow the original distribution \citep{Devroye1986}, as explained by Equation (\ref{eq:equ10}). 
%-----------------------------------------------------------
\begin{equation}
F^{-1}(u) \approx x_j \quad \forall\, u \in [0, 1], \quad \text{where } F_j \leq u < F_{j+1}
\label{eq:equ10}
\end{equation}
%-----------------------------------------------------------
Here, $F$ is CDF [Equation(\ref{eq:equ11})], $f$ is the normalized PDF [Equation (\ref{eq:equ12})], and $\Delta x$ is the step size [Equation (\ref{eq:equ13})].
%-----------------------------------------------------------
\begin{equation}
F_i = \sum_{k=1}^{i} f_k \Delta x, \quad F_N = 1
\label{eq:equ11}
\end{equation}
%-----------------------------------------------------------
\begin{equation}
f_i = \frac{f(x_i)}{\sum_{j=1}^{N} f(x_j) \, \Delta x}
\quad \text{so that} \quad
\sum_{i=1}^{N} f_i \, \Delta x = 1
\label{eq:equ12}
\end{equation}
%-----------------------------------------------------------
\begin{equation}
\Delta x = \frac{b - a}{N - 1} \quad \forall\, x \in [a, b], \quad \text{where } N \text{ is the total number of sample points}
\label{eq:equ13}
\end{equation}
%=============================================================
\\% -- Next Paragraph ----------------------------------------
\\% -- Add Gap -----------------------------------------------
\textbf{Inverse Transform Sampling - ALD: }The inverse transformation from the Asymmetric Laplace Distribution, producing skewness controlled by $\kappa$, is done by splitting the probability mass at $p=\kappa^2/(1+\kappa^2 )$, then using the piecewise inverse CDF on the left or right tail separately \citep{Kotz2001}, as shown by Equation (\ref{eq:equ14}):
%-----------------------------------------------------------
\begin{equation}
x =
\begin{cases}
m - \frac{1}{\lambda \kappa} \ln\left(\frac{1 - u}{1 - p}\right), & \text{if } u \geq p \\
m + \frac{\kappa}{\lambda} \ln\left(\frac{u}{p}\right), & \text{if } u < p
\end{cases}
\label{eq:equ14}
\end{equation}
%-----------------------------------------------------------
\\Here, 
\begin{flushleft}
$m \in \mathbb{R}$: Location parameter.\\[0.5em]
%-----------------------------------------------------------
$\lambda \in \mathbb{R}^+$: Scale parameter.\\[0.5em]
%-----------------------------------------------------------
$\kappa \in \mathbb{R}^+$: Asymmetry factor.\\[0.5em]
%-----------------------------------------------------------
$p = \frac{\kappa^2}{1 + \kappa^2}$: Skewness-related probability threshold.
%-----------------------------------------------------------
%\vspace{0.5em} % Optional spacing
\end{flushleft}
%============================================================================
\subsubsection{Real Data}\label{sec2_1_2}
%============================================================================
\begin{figure}[b]
\centering
\includegraphics[width=0.95\textwidth]{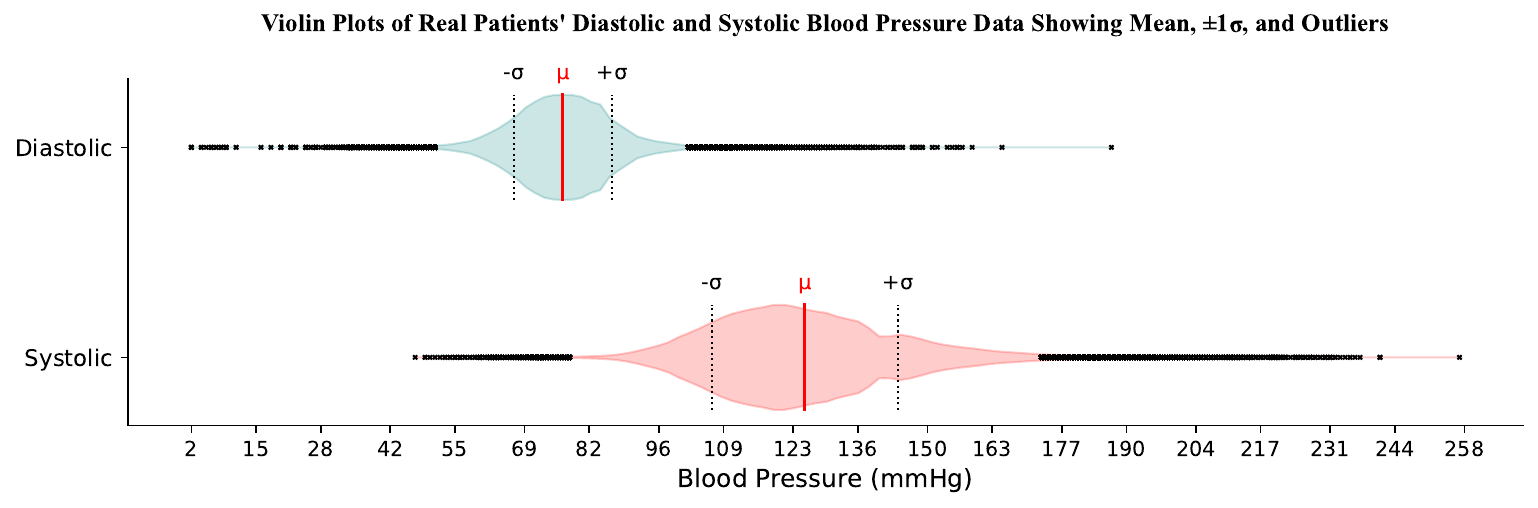}
\caption{Violin plots illustrating the distribution of real patients' systolic and diastolic blood pressure measurements. The plots display the mean ($\mu$), one standard deviation ($\pm 1\sigma$), and outliers beyond $\pm 2.5 \sigma$.}
\label{fig2}
\end{figure}
%***************************************************************
We validated our proposed DPR function using a real-world dataset comprising systolic and diastolic blood pressure measurements from 300,000 unique patients over a three-year period (2022-2024). We extracted the dataset from the Epic Electronic Health Record System \citep{Epic}. We masked all patients’ personal information to ensure data protection compliance.
\\% -- Next Paragraph -----------------------------------------
Between 2022 and 2024 (corresponding to the check-in period), we extracted three distinct datasets, each containing 100,000 unique patients, which were then concatenated to analyze a total of 300,000 data samples for each systolic and diastolic vital. We only focused on face-to-face patient visits within the specified query date range. Each patient ID was unique within a dataset, ensuring that every individual contributed only one pair of systolic and diastolic blood pressure readings per visit date. This approach maintained consistency and prevented duplication in the analysis. Fig.~\ref{fig2}  shows the violin plots of the diastolic and systolic blood pressure data used in the study, illustrating the overall distribution, including central tendency and variability.       
%****************************************************************************
\subsection{Probability Density Estimation}\label{sec2_2}
%****************************************************************************
We employed various methods for probability density estimation, including parametric approaches such as the Normal distribution and the Pearson Type I distribution, as well as non-parametric techniques like histograms, kernel density estimation (KDE), and our proposed DPR method, which are discussed in detail below.
%============================================================================
\subsubsection{Normal Distribution Estimation}\label{sec2_2_1}
%============================================================================
Normal distribution estimation is a fundamental parametric statistical technique used to model the distribution of a dataset. The normal distribution, also known as the Gaussian distribution, is characterized by its bell-shaped curve and is defined by two parameters: the mean ($\mu$) and the standard deviation ($\sigma$) \citep{Papoulis2002}. The probability density function (PDF) of a normal distribution is given by Equation (\ref{eq:equ15}):
%-----------------------------------------------------------
\begin{equation}
f(x) = \frac{1}{\sigma \sqrt{2\pi}} \exp\left( -\frac{(x - \mu)^2}{2\sigma^2} \right)
\label{eq:equ15}
\end{equation}
%-----------------------------------------------------------
Here,
\begin{flushleft}
$f(x)$: Probability density at point $x$.\\[0.5em]
$\mu$: Mean of the distribution.\\[0.5em]
$\sigma$: Standard deviation of the distribution.
\end{flushleft}
%============================================================================
\subsubsection{Pearson Type I Estimation}\label{sec2_2_2}
%============================================================================
The Pearson family of distributions is a versatile family characterized by the first four central moments (mean, variance, skewness, and kurtosis), which determine the shape and type of the distribution \citep{R2024, King2017,Yang2019}. First, we verified the Pearson distribution type by computing the Pearson discriminant value ($\kappa$) using the third- and fourth-order moments of the PDF, i.e., skewness and kurtosis parameters, via Equation (\ref{eq:equ16}). For all six synthetic datasets, we observed a negative Pearson discriminant value ($\kappa$), confirming the condition for Pearson Type-I, also described as the beta-I distribution, for which the PDF can be explained by Equation (\ref{eq:equ17}). To estimate the probability density for the 10000 evaluation points extracted from the lower and upper limits of the range with a constant step size, we utilized the \texttt{beta.pdf} function from the Python SciPy library \citep{Virtanen2020}. The shape parameters ($\delta_1,\ \delta_2$), location parameter ($\alpha$), and scale parameter ($\beta$) were estimated using the \texttt{beta.fit()} function.
%-----------------------------------------------------------
\begin{equation}
\kappa = \beta_2 - \frac{3}{2} \beta_1 - 3
\label{eq:equ16}
\end{equation}
%-----------------------------------------------------------
Here,
\begin{flushleft}
\(\beta_1 = \gamma_1^2 = \left( \frac{\mu_3}{\sigma^3} \right)^2\): Squared skewness.\\[0.5em]
%-----------------------------------------------------------
\(\beta_2 = \frac{\mu_4}{\sigma^4}\): Kurtosis.\\[0.5em]
%-----------------------------------------------------------
\(\mu_3\): $3^{rd}$ Central moment.\\[0.5em]
%-----------------------------------------------------------
\(\mu_4\): $4^{th}$ Central moment.\\[0.5em]
%-----------------------------------------------------------
\(\sigma\): Standard deviation.
\end{flushleft}
%-----------------------------------------------------------
\begin{equation}
f(x) = \frac{1}{B(\delta_1, \delta_2)} \left( \frac{x - \alpha}{\beta} \right)^{\delta_1 - 1} \left( 1 - \frac{x - \alpha}{\beta} \right)^{\delta_2 - 1}, \quad \text{for } \alpha \leq x \leq \alpha + \beta
\label{eq:equ17}
\end{equation}
%-----------------------------------------------------------
Here,
%-----------------------------------------------------------
\begin{flushleft}
\( B(\delta_1, \delta_2) \) is Beta function: 
\begin{equation}
 B(\delta_1, \delta_2) = \int_0^1 t^{\delta_1 - 1} (1 - t)^{\delta_2 - 1} \, dt = \frac{\Gamma(\delta_1)\Gamma(\delta_2)}{\Gamma(\delta_1 + \delta_2)}
\label{eq:equ18}
\end{equation}\\[0.5em]
%-----------------------------------------------------------
\(\delta_1\in \mathbb{R}^+\): Shape parameter — controls the distribution near the lower bound.\\[0.5em]
%-----------------------------------------------------------
\(\delta_2 \in \mathbb{R}^+\): Shape parameter — controls the distribution near the upper bound.\\[0.5em]
%-----------------------------------------------------------
\(\beta \in \mathbb{R}^+\): Scale parameter.\\[0.5em]
%-----------------------------------------------------------
\(\alpha \in \mathbb{R}\): Location parameter. 
\end{flushleft}
%============================================================================
\subsubsection{Tensor-Based Histogram Density Estimation (tHDE)}\label{sec2_2_3}
%============================================================================
To visualize the data distribution, we used density histograms generated via a custom TensorFlow function for Histogram Density Estimation (tHDE). If the number of bins (nBins) is not predefined in the tHDE function, then the number of bins (nBins) is estimated via Freedman–Diaconis Rule \citep{Freedman1981}, as shown in Equation (\ref{eq:equ19}). The function also applies a Gaussian filter, as described in Equation (\ref{eq:equ20}), to smooth the resulting density function. This smoothing uses a dynamic kernel size—no smaller than a predefined minimum threshold (default: 5)—based on the standard deviation ($\sigma$) to capture nearly the entire distribution ($\pm3\sigma$), along with a configurable smoothing parameter (default: 2.5). Both the minimum kernel size and smoothing parameter are adjustable via function arguments.
%-----------------------------------------------------------
\begin{equation}
\text{Number of Bins } (k) = \left\lceil \frac{\text{Range}}{h} \right\rceil = \left\lceil \frac{\max(X) - \min(X)}{\frac{2 \cdot \text{IQR}}{\sqrt[3]{N}}} \right\rceil
\label{eq:equ19}
\end{equation}
%-----------------------------------------------------------
Here,
%-----------------------------------------------------------
\begin{flushleft}
\( N \): Number of samples.\\[0.5em]
%-----------------------------------------------------------
\( \text{IQR} = Q_3 - Q_1 \): Interquartile Range.\\[0.5em]
%-----------------------------------------------------------
\( Q_1 \): 25th percentile ($1^{st}$ Quartile).\\[0.5em]
%-----------------------------------------------------------
\( Q_3 \): 75th percentile ($3^{rd}$ Quartile).\\[0.5em]
%-----------------------------------------------------------
\( h = 2 \cdot\text{IQR}/\sqrt[3]{N}\): Optimal bin width.\\[0.5em]
\end{flushleft}
%-----------------------------------------------------------
\begin{equation}
g_k = \frac{\exp\left( -\frac{1}{2} \left( \frac{k}{h} \right)^2 \right)}{\sum\limits_{j = -\left\lfloor m/2 \right\rfloor}^{\left\lfloor m/2 \right\rfloor} \exp\left( -\frac{1}{2} \left( \frac{j}{h} \right)^2 \right)}, \quad \text{for } k \in \left[ -\left\lfloor \frac{m}{2} \right\rfloor, \ldots, \left\lfloor \frac{m}{2} \right\rfloor \right]
\label{eq:equ20}
\end{equation}
%-----------------------------------------------------------
Here,
%-----------------------------------------------------------
\begin{flushleft}
\(\sum_k g_k = 1\): Normalization condition for the kernel weights.\\[0.5em]
%-----------------------------------------------------------
\(m = \max(\text{Size}_{\text{Th}}, \left\lfloor 6\sigma \right\rfloor)\): Kernel size.\\[0.5em]
%-----------------------------------------------------------
\(\text{Size}_{\text{Th}}\): Minimum kernel size threshold (default: 5).\\[0.5em]
%-----------------------------------------------------------
\(h\): Smoothing parameter (default: 2.5).%\\[0.5em]
\end{flushleft}
%============================================================================
\subsubsection{Kernel Density Estimation}\label{sec2_2_4}
%============================================================================
Kernel density estimation (KDE) is a statistically sound method for estimating a continuous distribution from a finite set of points, as KDE provides a non-parametric, noise-resilient view of the distribution while capturing both the central mass and the tails \citep{Zheng2013}. The estimated PDF using KDE utilizes the contributions of all data points scaled by a normalization factor and is explained by Equation (\ref{eq:equ21}) \citep{Silverman1998, Chan2017, Scott2015}:
%-----------------------------------------------------------
\begin{equation}
\tilde{f}(x) = \frac{1}{n h} \sum_{i=1}^{n} K\left( \frac{x - x_i}{h} \right)
\label{eq:equ21}
\end{equation}
%-----------------------------------------------------------
Here,
\begin{flushleft}
\( \tilde{f}(x) \): Estimated density at point \( x \).\\[0.5em]
\( n \): Number of data points.\\[0.5em]
\( h \): Bandwidth parameter.\\[0.5em]
\( K \): Kernel function.\\[0.5em]
\( x_i \): The \( i \)-th data point in the dataset.%\\[1.0em]
\end{flushleft}
\noindent We benchmarked Kernel Density Estimation (KDE) using two approaches:
\\% -- Next Paragraph -----------------------------------------
\\% -- Add Gap ------------------------------------------------
\textbf{\textit{(i) Python SciPy Kernel Density Estimation (SciPyKDE): }}As a baseline, we used the Gaussian KDE function from the SciPy library (SciPyKDE), which is built on NumPy \citep{Virtanen2020}. By default, this function estimates the bandwidth using Scott’s Rule \citep{Scott2015}.
\\% -- Next Paragraph -----------------------------------------
\\% -- Add Gap ------------------------------------------------
\textbf{\textit{(ii) Tensor-Based Kernel Density Estimation (tKDE): }}We implemented a custom KDE function using TensorFlow (tKDE), optimized for GPU acceleration to reduce computational time significantly. This tensor-based tKDE function estimates the probability density of 1D data using Gaussian kernels. It supports an optional bandwidth parameter, which defaults to Scott’s Rule for automatic selection. The function evaluates the PDF over either user-defined query points or an evenly spaced grid. It computes pairwise Gaussian kernel values, averages them to form the density estimate, and uses \verb|@tf.function(reduce_retracing=True)| for optimized graph execution and reduced retracing overhead, enabling fast and consistent performance.
\\% -- Next Paragraph -----------------------------------------
\\% -- Add Gap ------------------------------------------------
For consistency, we also approximated the bandwidth parameter in tKDE using Scott’s Rule \citep{Scott2015}, as described in Equation (\ref{eq:equ22}). Our optimized function (tKDE) significantly reduced computation time while maintaining almost the same accuracy metrics. While evaluating real patients’ systolic and diastolic datasets, we used the probability estimates generated by KDE methods (SciPyKDE and tKDE) as baselines for comparison with our proposed Dual Polynomial Regression (DPR) estimation method.
\\% -- Add Gap ------------------------------------------------
\begin{equation}
h = \sigma \cdot n^{-1/5}
\label{eq:equ22}
\end{equation}
\\% -- Add Gap ------------------------------------------------
However, a significant drawback of KDE is its substantial computational overhead \citep{Raykar2010,Chan2017}. The kernel function (K) and the difference function ($x-x_i$) require arrays shaped by the total number of data points (N). Consequently, estimating even a small window of the dataset demands considerable memory, leading to increased resource usage and computational time.
%============================================================================
\subsubsection{Dual Polynomial Regression (DPR) Estimation}\label{sec2_2_5}
%============================================================================
To robustly approximate the underlying probability density function (PDF) of unimodal data distributions, we employ a higher-order dual polynomial regression (DPR) approach. The process begins by constructing a smooth empirical estimate of the PDF using non-parametric density estimation—either tKDE or tHDE—then splitting the estimated PDF into two halves centered at the point of maximum probability (the peak). Each half is fitted with a separate polynomial regression, analogous to the concept used in ALD. This division is crucial as it allows the model to account for potential skewness in the distribution by fitting each side independently and capturing the asymmetry more precisely.
\\% -- Next Paragraph -----------------------------------------
Each polynomial is constructed via least-squares regression utilizing TensorFlow function (\texttt{tf.linalg.lstsq}) \citep{TensorFlow2024}, applied independently to the left and right halves of the estimated density function. We used the default fast computational method for \texttt{tf.linalg.lstsq} function, which solves the equations via Cholesky decomposition as explained by Equations (\ref{eq:equ24}) \citep{TensorFlow2024}.
\\% -- Add Gap ------------------------------------------------
\begin{equation}
\hat{y}_i = \sum_{j=0}^{P} c_{j\cdot}\, x_i^j
\label{eq:equ23}
\end{equation}
\\% -- Add Gap ------------------------------------------------
Where, C=$[c_0,c_1,\dots c_P]^T$ is vector of coefficients solved via  Cholesky decomposition :
\begin{equation}
C=(X^TX)^{-1}\cdot X^TY
\label{eq:equ24}
\end{equation}
Here, 
\begin{flushleft}
\( P \in \mathbb{Z}^+\): Polynomial order.\\[0.2em]
\( N \in \mathbb{Z}^+\): Number of samples.\\[0.2em]
\( C \in \mathbb{R}^{(P+1)\times1}\): Column vector of polynomial coefficients $[c_0,c_1, \dots c_P]^T$.\\[0.2em]
\( X \in \mathbb{R}^{N\times(P+1)}\): Vandermonde matrix built from data points.\\[0.2em]
\( Y \in \mathbb{R}^{N\times1}\): Column vector of observed outputs corresponding to each data point.\\%[0.5em]
\end{flushleft}
%-----------------------------------------------------------
Upon fitting, the two polynomial segments are concatenated to form a continuous representation of the complete probability density function (PDF). The return function is designed to operate on multi-dimensional tensors, assuming the last axis corresponds to the data points. This structure enables efficient parallel computation across batches, channels, or feature dimensions.
\\% -- Next Paragraph -----------------------------------------
To ensure shape stability, the start and end threshold points—defining the onset of the rising slope (left) and the termination of the falling slope (right)- are dynamically computed when not explicitly provided. This is accomplished by thresholding the full data sample using $\pm K$ times the standard deviation (default: K=5). Within these bounds, regions of rapid gradient change are identified using a gradient threshold scaled by a configurable Threshold Factor (default: 1\%). For the rising slope, all points below the start threshold, and for the falling slope, all points beyond the end threshold, are assigned a small positive minimum probability to suppress numerical instability. To further preserve numerical integrity and maintain the fidelity of the resulting distribution, all output values are clipped to a small positive minimum and constrained to remain below the peak of the density curve.
\\% -- Next Paragraph -----------------------------------------
The entire implementation is built in TensorFlow to leverage GPU acceleration and parallelism, except for the gradient thresholding module, which is implemented in NumPy for flexibility in array manipulation. The pseudocode for the Dual Polynomial Regression (DPR) is presented below (Algorithm~\ref{algo1}).

%----------------- Algorithm-1 | Part 1 -----------------
\begin{algorithm}[H]
\caption{Dual Polynomial Regression (DPR) }\label{algo1}
\begin{algorithmic}[ ]
\vspace{0.2em}  % <-- manual space 
\Require Input vector\ \( X \in \mathbb{R}^N \)
\vspace{0.2em}  % <-- manual space 
\Require Number of evaluation points for tKDE / tHDE\ \( J \in \mathbb{Z}^+ \)
\vspace{0.2em}  % <-- manual space 
\Require Polynomial order for curve fitting\ \( O \in \mathbb{Z}^+ \)
\vspace{0.2em}  % <-- manual space 
\Require Select return (default: T)\ \(F_{prob} \in \{T, F\}\ |\ T: Probabilty; F: PDF\)
\vspace{0.2em}  % <-- manual space 
\Require Range in units of $\sigma$ for trimming (default: 5)\ \( K \in \mathbb{R}^+ \)
\vspace{0.2em}  % <-- manual space 
\Require Scaling factor for gradient thresholding (default: 0.01)\ \(T_{factor} \in \mathbb{R}^+ \)
\vspace{0.2em}  % <-- manual space 
\Require Select backend (default: T)\ \(F_{KDE} \in \{T, F\}\ |\ T: tKDE;\ F: tHDE\)
\vspace{0.2em}  % <-- manual space 
\Require Min filter size for Gaussian smoothing (default: 3) \( fSize_{min} \in \mathbb{Z}^+ \)
\vspace{0.2em}  % <-- manual space 
\Statex \textbf{Optional} – Required for tHDE
\vspace{0.2em}  % <-- manual space 
\Require Smoothing parameter for Gaussian filter (default: 2.5) \( h \in \mathbb{Z}^+ \)
\vspace{0.2em}  % <-- manual space 
\Statex \textbf{Optional} – Required for tHDE
\vspace{0.2em}  % <-- manual space 
%-----------------------------------------------------------
\Ensure Estimated density function \( P(X) \)
\vspace{0.2em}  % <-- manual space 
%-----------------------------------------------------------
\Statex 1: Compute mean \( \mu \) and standard deviation \( \sigma \) of \( X \)
\vspace{0.2em}  % <-- manual space 
\Statex 2: Set thresholds: \( x_{\text{min}} = \mu - K\sigma \), \( x_{\text{max}} = \mu + K\sigma \)
\vspace{0.2em}  % <-- manual space 
\Statex 3: For the subset within the bounds estimate density using tKDE or tHDE:
%-----------------------------------------------------
\begin{equation}
f(\overleftrightarrow{x}) =
\begin{cases}
    \mathrm{tKDE}(\overleftrightarrow{x}), & F_{\text{KDE}} = \text{True} \\[0.5em]
    \mathrm{tHDE}(\overleftrightarrow{x}), & F_{\text{KDE}} = \text{False}
\end{cases}
\label{eq:equ25}
\end{equation}
%-----------------------------------------------------
\Statex 4: Sort and split density curve into two halves – Left and Right:
%-----------------------------------------------------
\begin{equation}
x_{\text{break}} = \arg\max_{x} f(\overleftrightarrow{x})
\label{eq:equ26}
\end{equation}
%-----------------------------------------------------
\Statex 5: Identify bounds where density gradient $<T_{\text{factor}}$\\
\vspace{0.2em}  % <-- manual space 
$\min(x_L)$: Bound for left half\\
\vspace{0.2em}  % <-- manual space 
$\max(x_R)$: Bound for right half\\
%-----------------------------------------------------
\begin{equation}
x_{\min} \;\leq\; \min(x_L) \;\leq\; \frac{x_{\min} + x_{\text{break}}}{2}
\label{eq:equ27}
\end{equation}
\Statex
\begin{equation}
\frac{x_{\text{break}} + x_{\max}}{2} \leq \max(x_R) \leq x_{\max}
\label{eq:equ28}
\end{equation}
%-----------------------------------------------------
% Manually set Counter
\vspace{0.2em}  % <-- manual space 
\Statex 6: Fit separate polynomials to left and right intervals utilizing least-square method: 
%-----------------------------------------------------
\begin{equation}
\hat{y}(x) =
\begin{cases}
\displaystyle \sum_{i=0}^{O} a_i x^i, & x \in [\min(x_L), x_{\text{break}}), \\[0.3em]
\displaystyle \sum_{i=0}^{O} b_i x^i, & x \in [x_{\text{break}}, \max(x_R)], \\[0.3em]
\epsilon, & \text{otherwise.}
\end{cases}
\label{eq:equ29}
\end{equation}
%-----------------------------------------------------
\end{algorithmic}
\end{algorithm}
%----------------- Algorithm-1 | Part 2 -----------------
\begin{algorithm}[H]
\addtocounter{algorithm}{-1}  % decrement the algorithm counter
\caption{Dual Polynomial Regression (DPR) – Continued ...}
\begin{algorithmic}[-1]
\vspace{0.2em}  % <-- manual space 
\Statex 7: Clip within bounds
\begin{equation}
\tilde{y}(x) =
\begin{cases}
\epsilon, & \hat{y}(x) < \epsilon, \\[0.2em]
y_{\max}, & \hat{y}(x) > y_{\max}=\hat{y}(x_{break}), \\[0.2em]
\hat{y}(x), & \text{otherwise.}
\end{cases}
\label{eq:equ30}
\end{equation}
%-----------------------------------------------------
\Statex 8: Normalize the PDF
\begin{equation}
p(x_k) = \frac{\tilde{y}_k}{\displaystyle \int_{x_{\min}}^{x_{\max}} \tilde{y}(x) \, dx}
\label{eq:equ31}
\end{equation}
%-----------------------------------------------------
\vspace{0.2em}  % <-- manual space 
\Statex \Return Normalized probability $p(x_k)$ if $F_{\text{prob}} = \text{True}$, else
\vspace{0.1em}  % <-- manual space 
\Statex \Return Callable PDF function to return $p(x_k)$
\vspace{0.1em}  % <-- manual space 
\Statex Here, $\epsilon$: Small constant (default: $10^{-12}$).
\vspace{0.1em}  % <-- manual space 
\end{algorithmic}
\end{algorithm}
%****************************************************************************
\vspace{-1.5em} % negative space removes gap
%****************************************************************************
\subsection{Comparative Analysis}\label{sec2_3}
%****************************************************************************
The various probability density estimation methods used in this study are benchmarked based on their accuracy in estimating the probability density, as well as the associated computational overhead. The Jensen-Shannon Divergence (JSD) is a method for measuring the similarity between two probability distributions. It is a symmetrized and smoothed version of the Kullback-Leibler (KL) divergence, which is always finite and symmetric, making it more stable and interpretable for comparing two PDFs from the same dataset \citep{Nielsen2019}. The Mean Squared Error (MSE) is a metric used to measure the average squared difference between predicted values and actual (true) values \citep{Dodge2008}. It measures the accuracy of predictions in relation to the actual data. The Pearson Correlation Coefficient ($\rho$) is a measure of the linear relationship between two continuous variables, which quantifies both the strength and direction of the linear association \citep{Dodge2008}. To quantify the accuracy of the estimation model, we measured the density estimation accuracy using MSE, JSD, and Pearson Correlation relative to the baseline PDF. Additionally, to ensure the validity of the estimated probability density functions, we verified that the Area Under the Curve (AUC) approximates 1 for each method when applied to the entire dataset, confirming that the estimated PDFs satisfy the normalization condition.
The analysis for accuracy and computational overhead across datasets for benchmarking is explained in detail below.
%============================================================================
\subsubsection{Computational Complexity}\label{sec2_3_1}
%============================================================================
We performed an asymptotic analysis to estimate the computational overhead of each function, considering both time and memory complexity. The results are summarized in Table~\ref{Table1}. Furthermore, we verified the execution time of each function call during probability estimation (training and inference) tasks and validated the theoretical assessment. 
%-----------------------------------------------------
\begin{table}[!t]
\caption{Theoretical time and space complexity of different probability density estimation methods.}
\label{Table1}
\centering
\scriptsize
\renewcommand{\arraystretch}{1.2}
\begin{tabularx}{\linewidth}{|P{1.1in}|P{1.1in}|P{1.0in}|X|}
\hline
\makecell[c]{\textbf{Function}} & 
\makecell[c]{\textbf{Time} \\ \textbf{Complexity}} & 
\makecell[c]{\textbf{Space} \\ \textbf{Complexity}} & 
\makecell[c]{\textbf{Remarks}}\rule{0pt}{4.0ex} \\
\hline
TensorFlow-Based tHDE & $O(N \log N+B)$ & $O(N + B)$ & Assuming $N \gg B$.\\ 
 %& & & N: Data points\\
 %& & & B: No. of Bins\\
 & & & Freedman–Diaconis Rule: $O(N \log N)$.\\
 & & & If $nBins$ is provided, reduces to: $O(N)$.\\
\hline
Normal Dist. & $O(M)$ & $O(M)$ & Linear function of evaluation points $M$. \\
\hline
Pearson Type I & $O(N + M)$ & $O(N + M)$ & \texttt{beta.fit} is $O(N)$. \\ 
 & & & Evaluating PDF for $M$ points is $O(M)$. \\
\hline
SciPyKDE\newline(SciPy Library) & $O(N \cdot M)$ & $O(N \cdot M)$ & Direct pairwise evaluation of kernel between $N$ data points and $M$ query points \citep{Gallego2022}.\\
\hline
TensorFlow-Based tKDE & $O(N \cdot M)$ & $O(N \cdot M)$ & Memory grows with $(N, M)$ matrix. \\ 
 & & & Batch dimensions scale linearly in memory, while time remains nearly constant due to parallel tensor operations. \\
\hline
DPR using tKDE\newline(Training)  & $O(N \cdot M + M \cdot O^2)$ & $O(N \cdot M + M \cdot O)$ & Assuming $N \gg M \cdot O$. \\ 
 & & & Kernel evaluation dominates. \\
\hline
DPR using tHDE\newline(Training) & $O(N \log N + M \cdot O^2)$ & $O(N + M \cdot O)$ & Sorting dominates for $N \gg M$. \\ 
 & & & Polynomial fitting dominates for large $O$. \\
\hline
DPR-tKDE/tHDE\newline(Evaluation)  & $O(M \cdot O)$ & $O(M \cdot O)$ & Polynomial evaluation of order $O$ dominates. \newline Batch dimensions scale linearly in memory; time nearly constant due to tensor parallelism. \\
\hline
\end{tabularx}
\footnotetext{N: Total data points.}
\footnotetext{B: Total bins used for Histogram.}
\footnotetext{M: Evaluation points.}
\footnotetext{O: Polynomial order used for regression.}
\end{table}
%***************************************************************
%***************************************************************
\begin{figure}[t]
\centering
\includegraphics[width=\textwidth]{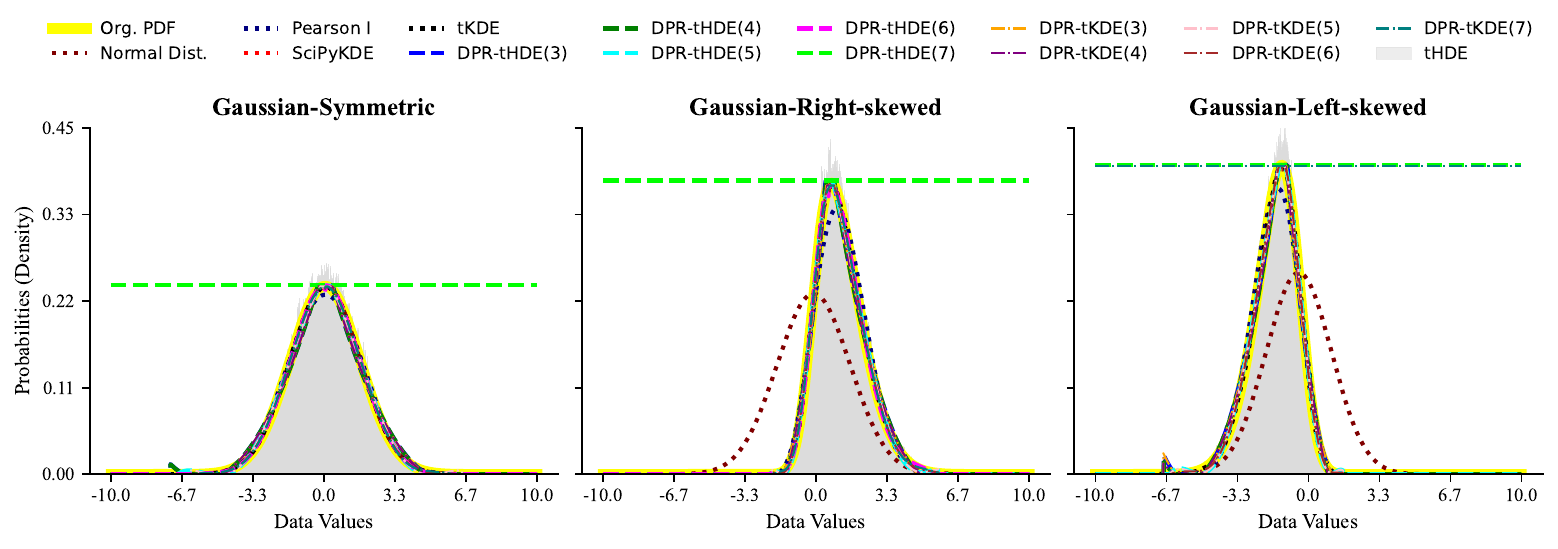}
\caption{Comparison of probability density estimation methods for synthetic data generated for Gaussian distributions with and without skewness. Each subplot shows the ground truth PDF in yellow, along with estimated PDFs from different approaches. Saturation in DPR at higher orders was caused by numerical instability within \texttt{tf.linalg.lstsq}.}
\label{fig3}
\end{figure}
%***************************************************************
\begin{figure}[t]
\centering
\includegraphics[width=\textwidth]{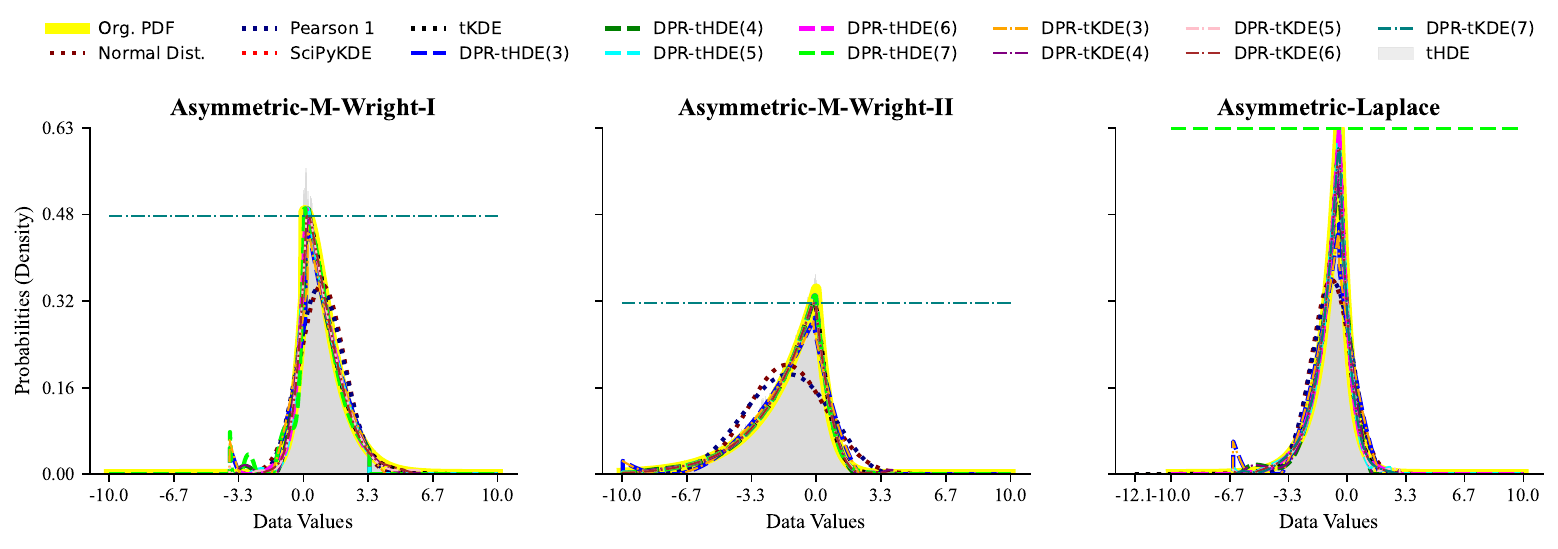}
\caption{Comparison of probability density estimation methods for synthetic data generated for AMW-I, AMW-II, and ALD. Each subplot shows the ground truth PDF in yellow, along with estimated PDFs from different methods. Saturation in DPR at higher orders was caused by numerical instability within \texttt{tf.linalg.lstsq}}
\label{fig4}
\end{figure}
%============================================================================
\subsubsection{Synthetic Data Evaluation}\label{sec2_3_2}
\begin{figure}[t]
\centering
\includegraphics[width=\textwidth]{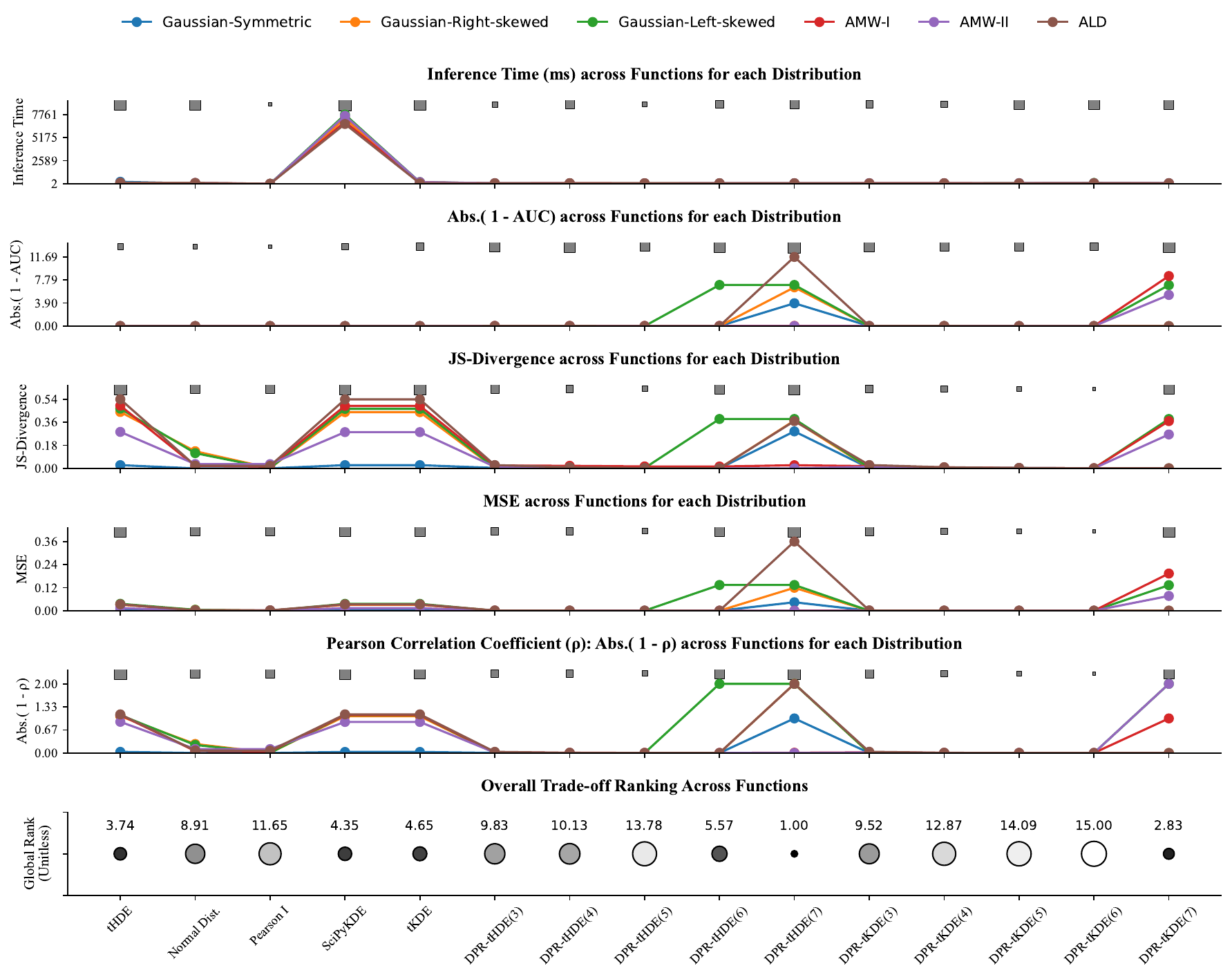}
\caption{Comparison of probability density estimation functions across synthetic datasets. Subplots show Inference Time, AUC, JSD, MSE, and Pearson correlation coefficient — AUC and correlation use $|1-value|$. Non-finite values were clipped to twice the maximum finite value. Squares above each plot indicate mean-performance ranks, with smaller squares denoting better ranks. The last subplot shows the overall trade-off ranking across all metrics.}
\label{fig5}
\end{figure}

%============================================================================
For the synthetic data generated for all six distributions as discussed above in Section~\ref{sec2_1_1} \textbf{`Synthetic Data'}, we estimated the probability density over 10,000 evaluation points, representing 5\% of the total generated data points. We benchmarked several methods, including parametric and non-parametric estimations, as discussed in Section~\ref{sec2_2} \textbf{`Probability Density Estimation'}. For parametric estimation, we considered the Normal distribution and the Pearson Type I distribution estimations. Non-parametric estimations included the Histogram utilizing tHDE, Kernel Density estimation via SciPyKDE and tKDE, along with our proposed Dual Polynomial Regression (DPR) estimation method with polynomial orders ranging from 3 to 7. 
\\
We verified accuracy via computing the Mean Squared Error (MSE), Jensen-Shannon Divergence (JSD), and Pearson correlation coefficient ($\rho$) with respect to the baseline PDF generated via the distribution function as a start point. Fig.~\ref{fig3} illustrates the density estimation via different methods used for the variants of the Gaussian distribution. AMW-I, AMW-II, and ALD are illustrated in Fig.~\ref{fig4}. A consolidated comparative analysis is presented in Fig.~\ref{fig5}, which shows the overall ranking as well as the individual rankings based on inference time, normality, and accuracy metrics. 
\\
For DPR, we observed numerical instability in Cholesky decomposition within \texttt{tf.linalg.lstsq} at higher orders, due to the inherent sensitivity of Cholesky decomposition to ill-conditioned matrices, which can lead to computational breakdowns \citep{Fukaya2018}. Exploring alternative or improved approaches to the Cholesky decomposition for achieving stable output in higher orders is left for future work.
%***************************************************************

%============================================================================
\subsubsection{Real Data Evaluation (Systolic \& Diastolic)}\label{sec2_3_3}
%============================================================================
\begin{figure}[b]
\centering
\includegraphics[width=0.9\textwidth]{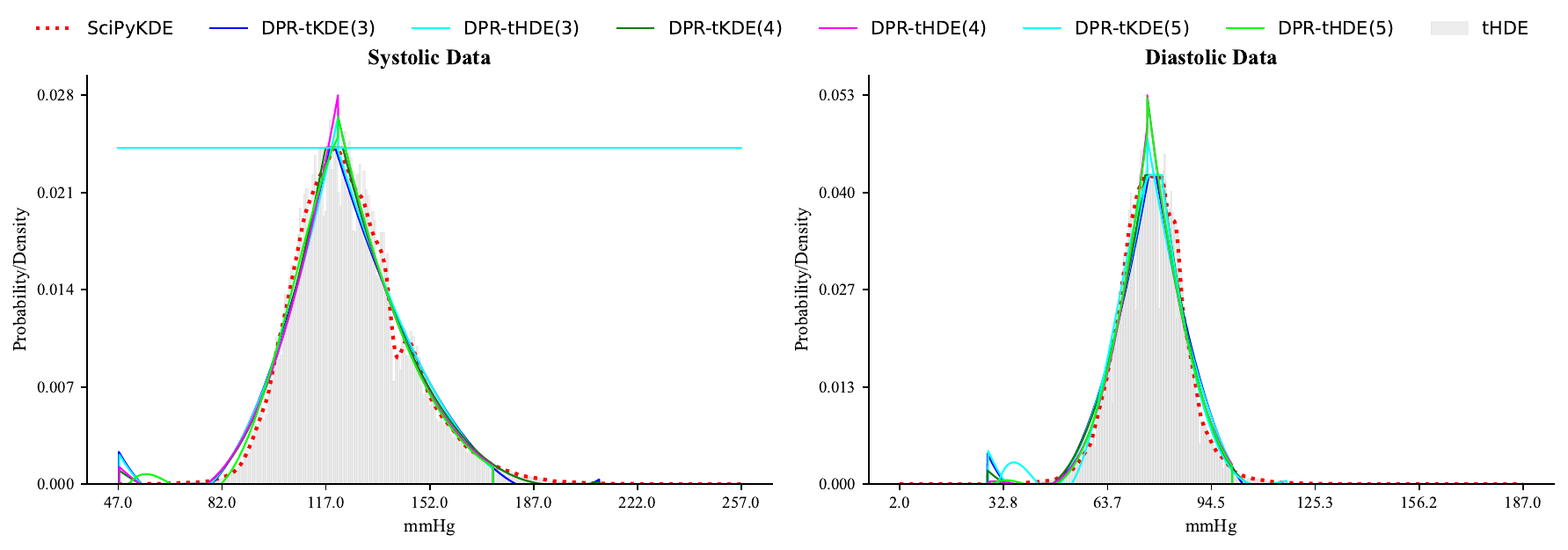}
\caption{Estimated probability density functions for real-world systolic (left) and diastolic (right) blood pressure data using various methods. SciPyKDE (red dotted line) serves as the baseline. DPR-based estimations are shown for both tKDE and tHDE variants with polynomial orders 3 to 5. Histograms via tHDE are visualized as gray-shaded areas, with bin counts determined using the Freedman–Diaconis rule. DPR using tKDE with Order 5 shows saturation for systolic data due to numerical instability in Cholesky decomposition.}
\label{fig6}
\end{figure}
\begin{figure}[t]
\centering
\includegraphics[width=\textwidth]{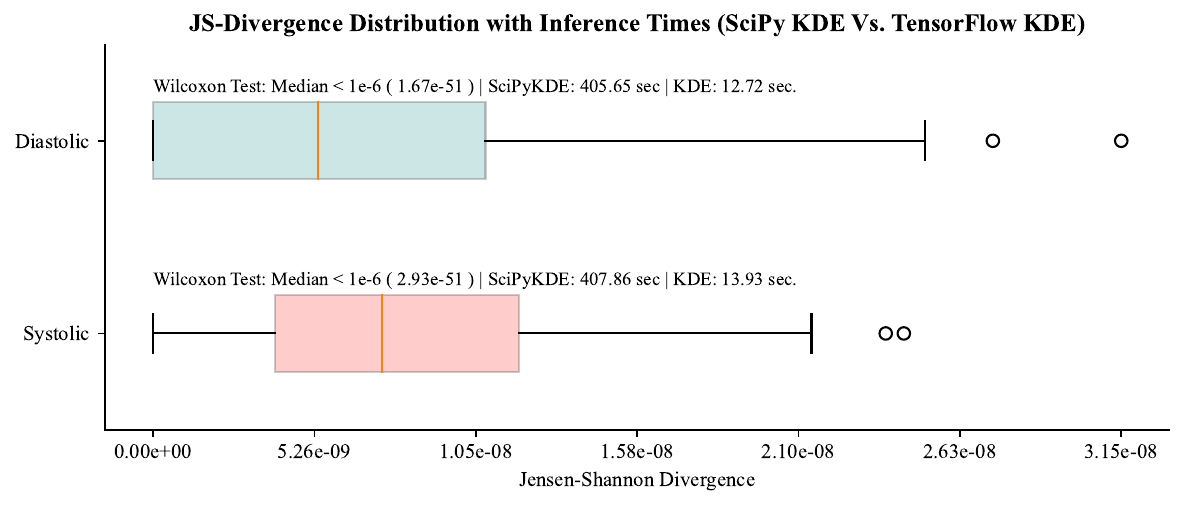}
\caption{Boxplots illustrating JS-Divergence (JSD) across 300 sliding windows for SciPyKDE and TensorFlow-based tKDE on systolic and diastolic data. P-values from the Wilcoxon signed-rank test, assessing whether the median JSD is significantly below $1\times10^{-6}$, are shown in the plot. Evaluation times for both methods are included to facilitate a comparison of computational efficiency.}
\label{fig7}
\end{figure}
\begin{figure}[t]
\centering
\includegraphics[width=\textwidth]{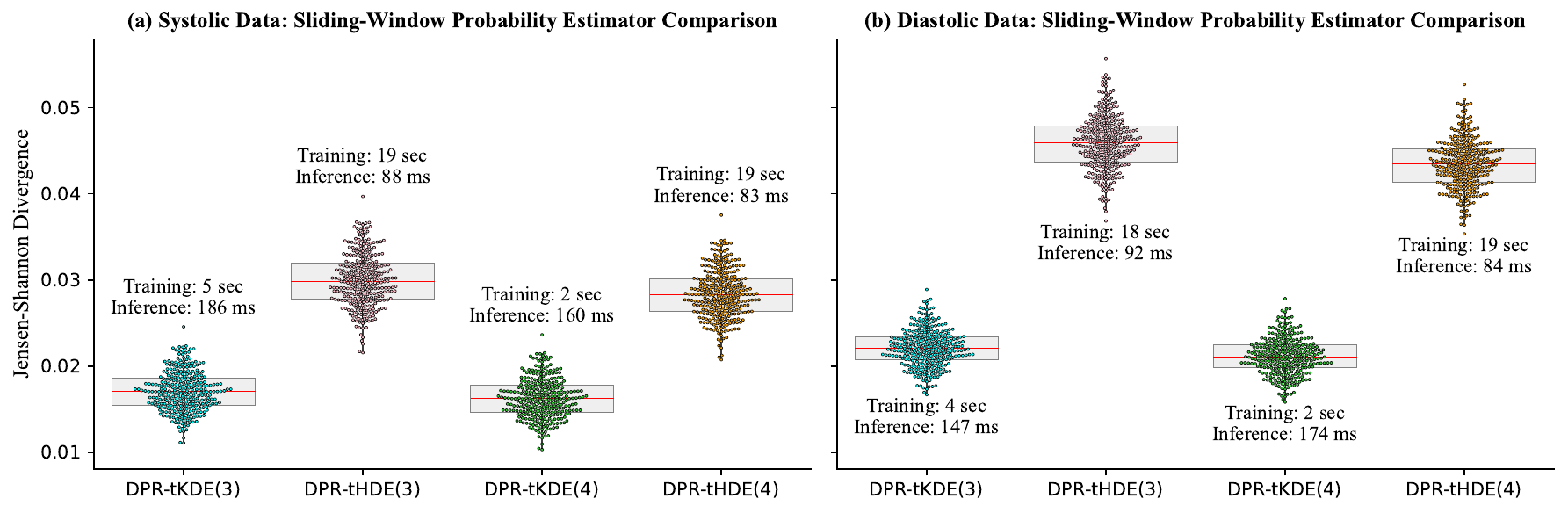}
\caption{Box plots with overlaid swarm plots showing Jensen-Shannon Divergence (JSD) across 300 sliding windows for systolic (a) and diastolic (b) data. DPR-based estimators using tKDE and tHDE backends with polynomial orders 3 and 4 are compared against the TensorFlow-based tKDE baseline. Execution times for each method are annotated above the respective distributions to highlight computational efficiency.}
\label{fig8}
\end{figure}

%***************************************************************
For real-world systolic and diastolic data, we estimated the probability density over 15,000 evaluation points, representing 5\% of the total 300,000 data points. To visualize the distribution, we used tHDE, where we estimated the number of bins using the Freedman-Diaconis Rule. Here, we used the kernel density estimation from the Python' SciPy library (SciPyKDE) as the baseline, since the custom TensorFlow KDE function (tKDE) was unable to complete due to excessive memory usage ($\sim18$ GB) when processing 300,000 data points over 15,000 evaluation points, leading to GPU memory saturation and kernel termination. 
\\% -- Next Paragraph -----------------------------------------
With reference to the baseline probability density estimated via SciPyKDE, we evaluated the probability density estimations employing the proposed DPR function with orders 3-5 for 15000 evaluation points spanning the actual data range. We limited the order to a maximum of 5 to prevent numerical instability caused by Cholesky decomposition within \texttt{tf.linalg.lstsq}. To restrict memory utilization while training the DPR models, utilizing 300,000 samples of systolic and diastolic data via tKDE, we used smaller evaluation points: 1,000 evaluation points. While training DPR via tHDE, we used 293 and 477 bins, estimation based on the Freedman-Diaconis Rule.
\\% -- Next Paragraph -----------------------------------------
We used the same metrics — JSD, MSE, and Pearson correlation coefficient — to validate the accuracy, in addition to training and inference times. Fig.~\ref{fig6} illustrates the density estimation via different methods used, and Table~\ref{Table2} and Table~\ref{Table3} show the metrics for the Systolic and Diastolic datasets, respectively.
\\% -- Next Paragraph -----------------------------------------
To simulate probability estimations for real-life scenarios of predictive modeling, we also performed localized probability estimation using a sliding window approach consisting of 300 windows, each containing 1,000 samples. The methods evaluated included Kernel density estimation from the Python’ SciPy library (SciPyKDE), the custom TensorFlow KDE function (tKDE), and DPR using both tKDE and tHDE, with polynomial orders ranging from 3 to 4. For the sliding window, we excluded the polynomial order 5, as in the previous case, we observed numerical instability encountered for the 5th order during Cholesky decomposition. 
\\% -- Next Paragraph ----------------------------------------
\\
%**************************************************************
\begin{table}[t]
\centering
\caption{Comparison of probability density estimation methods for systolic dataset.}
\label{Table2}
%\scriptsize
\renewcommand{\arraystretch}{1.2}
%\resizebox{\textwidth}{!}{%
\begin{tabular}{|l|c|c|c|c|c|c|}
\hline
\textbf{Function}\rule{0pt}{4.5ex} & 
\makecell{\textbf{Training}\\\textbf{Time (ms)}}\rule{0pt}{4.5ex} & 
\makecell{\textbf{Inference}\\\textbf{Time (ms)}}\rule{0pt}{4.5ex} & 
\textbf{AUC}\rule{0pt}{4.5ex} & 
\textbf{JSD}\rule{0pt}{4.5ex} & 
\textbf{MSE}\rule{0pt}{4.5ex} & 
\makecell{\textbf{Pearson}\\\textbf{Corr. Coef.}}\rule{0pt}{4.5ex} \\
\hline
HDE & -- & 17144 & 1.00 & -- & -- & -- \\
\textbf{SciPyKDE} & -- & \textbf{15290} & \textbf{1.00} & -- & -- & -- \\
DPR-tKDE(3) & 1112 & 67 & 0.99 & 0.010 & 8.5e-07 & 0.99 \\
DPR-tHDE(3) & 18239 & 68 & 0.99 & 0.012 & 9.2e-07 & 0.99 \\
DPR-tKDE(4) & 1203 & 99 & 1.00 & 0.005 & 5.7e-07 & 1.00 \\
DPR-tHDE(4) & 18250 & 69 & 0.99 & 0.010 & 8.0e-07 & 0.99 \\
DPR-tKDE(5) & 1223 & 82 & 5.05 & 0.273 & 4.3e-04 & -- \\
DPR-tHDE(5) & 18329 & 69 & 0.98 & 0.010 & 4.4e-07 & 1.00 \\
\hline
\end{tabular}%
\footnotetext{Baseline density estimation is performed using SciPyKDE. DPR with tKDE or tHDE backends is compared across polynomial orders (shown in parentheses) using JSD, MSE, and Pearson correlation coefficient as accuracy metrics.}
\end{table}
%***************************************************************
%\vspace{-1.5em}  % adjust value as needed
\begin{table}[t]
\caption{Comparison of probability density estimation methods for the diastolic dataset.}
\label{Table3}
\centering
%\scriptsize
\renewcommand{\arraystretch}{1.2}
\begin{tabular}{|l|c|c|c|c|c|c|}
\hline
\textbf{Function}\rule{0pt}{4.5ex} & 
\makecell{\textbf{Training}\\\textbf{Time (ms)}}\rule{0pt}{4.5ex} & 
\makecell{\textbf{Inference}\\\textbf{Time (ms)}}\rule{0pt}{4.5ex} & 
\textbf{AUC}\rule{0pt}{4.5ex} & 
\textbf{JSD}\rule{0pt}{4.5ex} & 
\textbf{MSE}\rule{0pt}{4.5ex} & 
\makecell{\textbf{Pearson}\\\textbf{Corr. Coef.}}\rule{0pt}{4.5ex} \\
\hline
HDE & -- & 17137 & 1.00 & -- & -- & -- \\
\textbf{SciPyKDE} & -- & \textbf{14031} & \textbf{1.00} & -- & -- & -- \\
DPR-tKDE(3) & 1151 & 68 & 0.99 & 0.013 & 3.1e-06 & 0.99 \\
DPR-tHDE(3) & 18233 & 70 & 0.98 & 0.015 & 3.4e-06 & 0.99 \\
DPR-tKDE(4) & 1238 & 71 & 0.99 & 0.008 & 1.7e-06 & 0.99 \\
DPR-tHDE(4) & 18200 & 70 & 0.99 & 0.009 & 2.7e-06 & 0.99 \\
DPR-tKDE(5) & 1277 & 70 & 1.00 & 0.013 & 1.3e-06 & 1.00 \\
DPR-tHDE(5) & 18230 & 71 & 0.99 & 0.009 & 2.4e-06 & 0.99 \\
\hline
\end{tabular}
\footnotetext{Baseline density estimation is performed using SciPyKDE. DPR with tKDE or tHDE backends is compared across polynomial orders (shown in parentheses) using JSD, MSE, and Pearson correlation coefficient as accuracy metrics.}
\end{table}
%**************************************************************
\\
\noindent For the sliding windows, we accessed the JSD computed using density estimation via SciPyKDE and TensorFlow-based tKDE for both Systolic and Diastolic data (Fig.~\ref{fig7}). A one-sided Wilcoxon signed-rank test was performed to assess whether the median JSD is significantly lower than a small threshold ($1\times10^{-6}$), with the null hypothesis $H_0$: median(JSD) $\geq1\times10^{-6}$ and the alternative hypothesis $H_1$: median(JSD) $<1\times10^{-6}$, indicating near-perfect similarity between the estimated and reference distributions. Also, we verified the required evaluation time for all 300 windows via both methods.
\\% -- Next Paragraph -----------------------------------------
We used the tKDE as the baseline and computed the JSD for each window to evaluate the DPR-based estimations against the tKDE reference. We analyzed the variability and median trends by plotting box plots with overlaid swarm plots of the JSD for each window and across different methods of DPR utilized (Fig.~\ref{fig8}). Additionally, we verified the training and inference times for each estimator to compare computational cost. Also performed a one-sided Mann–Whitney U test to compare the JSD distributions between polynomial orders 3 and 4 for both internal function tKDE and tHDE for both systolic and diastolic data to understand the impact of order. 

%***************************************************************
%------------------------------------------------------

%****************************************************************************
\subsection{Software and Hardware Environment}\label{sec2_4}
%****************************************************************************
All codes were developed in Python (v3.8.12) using NumPy (v1.22.3) and Keras-TensorFlow (v2.11.0), with the SciPy library (v1.10.0) employed for statistical and numerical computations. All computations were performed on a MacBook Pro with an Apple M1 Max chip (10-core CPU: 8 performance cores and 2 efficiency cores), one integrated GPU, and 64 GB of unified memory.
To ensure reproducibility, random number generation was controlled by explicitly setting seeds. All computational overhead metrics reported in this paper are based on this specific hardware and software configuration.
%***************************************************************
%@@@@@@@@@@@@@@@@@@@@@@@@@@@@@@@@@@@@@@@@@@@@@@@@@@@@@@@@@@@@@@@@@@@@@@@@@@@@
\section{Results}\label{sec3}
%@@@@@@@@@@@@@@@@@@@@@@@@@@@@@@@@@@@@@@@@@@@@@@@@@@@@@@@@@@@@@@@@@@@@@@@@@@@@
At first, we assessed the performance of various probability density estimation methods on synthetic datasets derived from six distinct unimodal distributions that replicate common statistical properties observed in real-world scenarios, including skewness and asymmetry. Fig.~\ref{fig3} illustrates the comparison of the baseline PDF and the estimated PDFs via different techniques for various Gaussian distributions, including both symmetric and asymmetric distributions. Whereas Fig.~\ref{fig4} illustrates a similar comparison of PDFs generated via different techniques for asymmetrical distributions: AMW-I, AMW-II, and ALD.
\\% -- Next Paragraph -----------------------------------------
Fig.~\ref{fig5} presents a comparative analysis of various metrics—Inference Time, Area Under the Curve (AUC), Jensen-Shannon Divergence (JSD), Mean Squared Error (MSE), and Pearson correlation coefficient—plotted for each function across all distributions. It also shows the overall trade-off ranking, aggregating performance across all metrics to highlight the balance between accuracy and computational efficiency among methods such as tHDE, SciPyKDE, tKDE, and DPR variants. To compute the overall trade-off ranking (also referred to as the global ranking), we assigned a weight of 0.5 to inference time and 1.0 to all other metrics. The final global ranking was normalized on a scale from 1 to 15. Fig.~\ref{fig5} indicates that the DPR with tHDE and tKDE in the backend, using 7th order, resulted in saturated estimations mainly caused by numerical instability encountered in least square estimation. DPR with tHDE in the backend, using 6th order, also resulted in saturated estimations for a Gaussian left-skewed distribution. Ignoring the 6th and 7th orders, we observed that DPR with tKDE and order 5 outperformed, followed by DPR with tHDE and order 5. It is evident that in the DPR, using both tKDE and tHDE in the backend, the global ranking decreases with the order utilized during polynomial regression. The parametric Pearson Type I function also shows a good global ranking ($\sim11.6$) - however, as we observed the plots in Fig.~\ref{fig3} and Fig.~\ref{fig4}, we noticed Pearson Type I captures the shape of the distribution well for Gaussian types (with or without skewness) but fails to capture the asymmetrical distributions, such as AMW-I, AMW-II, and ALD.
\\% -- Next Paragraph -----------------------------------------
In Fig.~\ref{fig2}, the violin plots for real patients’ systolic and diastolic blood pressure data clearly indicate that the distributions are not truly Gaussian. In particular, the systolic blood pressure data exhibit a long-tailed, asymmetric distribution, deviating significantly from the bell-shaped curve as defined by the normal distribution. This observation underscores the importance of using flexible, non-parametric density estimation techniques. 
\\% -- Next Paragraph -----------------------------------------
Similar to synthetic datasets, for the systolic and diastolic blood pressure dataset, we considered 5\% of the total dataset size as evaluation points, i.e., 15000 points, expanding within the range with a constant step size. SciPyKDE was used as the baseline, as the TensorFlow-based tKDE failed due to kernel termination, caused by memory usage exceeding $\sim18$ GB. 
DPR model variants were trained using 300,000 samples with reduced evaluation points (1,000 for tKDE and bin counts of 293 and 477 for tHDE, based on the Freedman–Diaconis rule). During the training of DPR, we limited the polynomial orders to 3 - 5 to avoid saturation due to numerical instability caused by orders $> 5$, as observed in the case of synthetic datasets.
\\% -- Next Paragraph -----------------------------------------
Table~\ref{Table2} and Table~\ref{Table3} present a detailed comparison of probability density function estimation methods applied to systolic and diastolic blood pressure data, respectively. Each table reports training and inference times (in milliseconds), along with key accuracy metrics — AUC, JSD, MSE, and P Pearson correlation coefficient — to evaluate the performance of each method. The proposed Dual Polynomial Regression (DPR) approach is shown with varying polynomial orders (O), highlighting its adaptability and trade-offs across computational efficiency and estimation accuracy. It shows that the DPR inference time is substantially lower than the SciPyKDE implementation, corresponding to a reduction of more than 99\% in execution time. We also observe that DPR with tKDE as the backend requires more than 90\% less training time compared to DPR with tHDE as the backend.
\\% -- Next Paragraph -----------------------------------------
Sliding-window analysis was performed using 300 windows of 1,000 samples each for both systolic and diastolic datasets. We first analyzed the JSD  between the density estimations via SciPyKDE and tKDE. A one-sided Wilcoxon signed-rank test was performed to verify whether the median JSD is significantly lower than a small threshold value ($1\times10^{-6}$), indicating near-perfect similarity between the estimated and reference distributions. Fig.~\ref{fig7} illustrates the box plots of the JSD across 300 sliding windows computed for systolic and diastolic data, including Wilcoxon test p-values for the null hypothesis that the median JSD is $\geq1\times10^{-6}$, along with evaluation times to compare computational efficiency. From the resulting p-values, we can reject the null hypothesis. We observed that tKDE is approximately $30\times$ faster than SciPyKDE, as it is GPU optimized and performs computations on a batch, processing all 300 windows in parallel.
\\% -- Next Paragraph -----------------------------------------
For sliding window analysis, utilizing DPR-based methods, we considered polynomial orders 3 and 4 only. For analyzing the DPR-based method with tHDE and tKDE as backends, we used the density estimations of the 300 windows (both systolic and diastolic data) via TensorFlow-based tKDE as a baseline. Accuracy was evaluated using JSD and visualized in Fig.~\ref{fig8} via box plots with overlaid swarm plots. Each plot is also annotated with training and inference times, enabling a joint assessment of accuracy via JSD and computational cost, thereby providing insight into the trade-offs between tKDE- and tHDE-based approaches. The mean inference time for estimating the density of 300 windows, measured across all 8 cases, is significantly lower ($\approx126 \pm 41$ ms), making this approach practically feasible for real-time applications.
\\% -- Next Paragraph -----------------------------------------
We observed that the actual training time for the DPR function with tKDE as backend is approximately $3\times$ faster than with tHDE as backend, which corresponds to the ratio of the theoretical polynomial operations. For tKDE, we used 100 evaluation points, whereas for tHDE, we used 293 and 477 evaluation points (based on the Freedman-Diaconis rule) for the systolic and diastolic datasets, respectively. Considering the sliding-window analysis with 4th-order polynomial regression on the systolic dataset (M = 293 for tHDE), the theoretical polynomial operations are:
%-----------------------------------------------------
\begin{flushleft}
DPR (tKDE, Order 4): $T_1 = M \cdot O^2 = 100 \cdot 16 = 1,600 \approx 0.0016$ million operations \\[0.5em]
DPR (tHDE, Order 4): $T_2 = M \cdot O^2 = 293 \cdot 16 = 4,688 \approx 0.0047$ million operations \\[0.5em]
Ratio: $T_2:T_1 \approx 3$%\\[0.5em]
\end{flushleft}
%-----------------------------------------------------
\noindent The results show that DPR with tKDE as backend yields the lowest JSD values and significantly reduced training time. We performed one-sided Mann–Whitney U tests ($\alpha = 0.01$) to compare JSD distributions between polynomial orders 3 and 4. The null hypothesis ($H_0$: Order 4 does not have a significantly lower median JSD than Order 3) was rejected for both systolic and diastolic datasets, for both tKDE- and tHDE-based backends, indicating that the fourth order exhibits a significantly lower median JSD than the third order in all cases. 
%@@@@@@@@@@@@@@@@@@@@@@@@@@@@@@@@@@@@@@@@@@@@@@@@@@@@@@@@@@@@@@@@@@@@@@@@@@@@
\section{Conclusion}\label{sec4}
%@@@@@@@@@@@@@@@@@@@@@@@@@@@@@@@@@@@@@@@@@@@@@@@@@@@@@@@@@@@@@@@@@@@@@@@@@@@@
This study presents a comprehensive evaluation of various probability density estimation techniques across both synthetic and real-world datasets, with a particular focus on the proposed Dual Polynomial Regression (DPR) framework. Our findings demonstrate that DPR, when paired with TensorFlow-based tKDE or tHDE backends, offers a compelling balance between computational efficiency and estimation accuracy.
\\% -- Next Paragraph -----------------------------------------
On synthetic datasets, for the DPR method using both tKDE and tHDE as backends, the global ranking increases with the polynomial order used during regression. However, it shows saturation beyond a particular order due to numerical instability—specifically in the Cholesky decomposition during regression fitting. It was observed that using polynomial orders $\leq5$ limits the risk of saturation, and the DPR variants, particularly with tKDE as the backend, outperformed classical approaches. These models achieved the highest global rankings across key accuracy metrics, including AUC, JSD, MSE, and Pearson Correlation, while maintaining significantly lower inference times.
\\% -- Next Paragraph -----------------------------------------
For real-world blood pressure data, the non-Gaussian nature of the distributions—particularly the long-tailed asymmetry in systolic measurements—highlighted the limitations of traditional parametric methods and underscored the need for flexible, non-parametric approaches. DPR variants demonstrated  $>99\%$ reduction in inference time compared to SciPyKDE, maintaining nearly the same accuracy. Sliding-window analysis further showed that the TensorFlow-based tKDE function is $\times30$ times faster compared to SciPyKDE. In terms of inference time, DPR methods are significantly faster, able to estimate a batch of 300 windows within just hundreds of milliseconds, making them suitable for real-time applications. DPR methods utilizing 4th-order polynomial regression consistently yielded lower JSD values than 3rd-order polynomial regression, as confirmed by Mann–Whitney U tests. Additionally, DPR variants with tKDE as a backend outperformed those with tHDE.
\\% -- Next Paragraph -----------------------------------------
Where the data size is within computational limits, DPR with tKDE as the backend is recommended to achieve better accuracy. Since DPR using tHDE requires less memory for training compared to DPR with tKDE, tHDE as the backend should be preferred for significantly large datasets to avoid termination due to memory overuse. Hence, DPR with tHDE is a more suitable choice in memory-constrained environments, especially when working with high-volume data.
\\% -- Next Paragraph -----------------------------------------
Overall, the DPR framework proves to be a scalable and accurate solution for unimodal density estimation, particularly well-suited for large-scale applications, especially under non-Gaussian and data-slicing conditions, and computationally efficient for real-time applications. Future work may explore alternative or improved approaches to the Cholesky decomposition for achieving stable output with higher orders. Future work could further explore adaptive order selection and hybrid backend strategies to enhance performance and improve generalizability across a broader range of data modalities.
\\% --add Line -----------------------------------------
\begin{comment}
%@@@@@@@@@@@@@@@@@@@@@@@@@@@@@@@@@@@@@@@@@@@@@@@@@@@@@@@@@@@@@@@@@@
\section*{Code Availability}\label{sec5}
%@@@@@@@@@@@@@@@@@@@@@@@@@@@@@@@@@@@@@@@@@@@@@@@@@@@@@@@@@@@@@@@@@@
The code used in this study is publicly available at 
\href{https://github.com/ShanSarkar75/DPR/}{https://github.com/ShanSarkar75/DPR/}. 
The repository includes all scripts, datasets (where permitted by data sharing policies), and environment 
specification files (including \texttt{requirements.txt}) to facilitate full reproduction of the analyses and figures. 
In addition, the implemented Python library functions are available as an installable package via the Python Package Index (PyPI): 
\href{https://pypi.org/project/estimatePDF/}{https://pypi.org/project/estimatePDF/}.
%******************************************************
\bmhead{Acknowledgements}
The authors gratefully acknowledge the support of colleagues and institutions that contributed to this research.

\end{comment}

%@@@@@@@@@@@@@@@@@@@@@@@@@@@@@@@@@@@@@@@@@@@@@@@@@@@@@@@@@@@@@@@@@@
\section*{Declarations}\label{sec6}
%@@@@@@@@@@@@@@@@@@@@@@@@@@@@@@@@@@@@@@@@@@@@@@@@@@@@@@@@@@@@@@@@@@
\begin{itemize}
\item The authors declare no competing interests.
\item \textbf{Funding: } No funds, grants, or other support was received.
% Conflict of interest/Competing interests (check journal-specific guidelines for which heading to use)
%\item Ethics approval and consent to participate
%\item Consent for publication
%\item Data availability 
\item \textbf{Code availability:} The code used in this study is publicly available at \href{https://github.com/ShanSarkar75/DPR/}{https://github.com/ShanSarkar75/DPR/}. 
The repository includes all scripts, datasets (where permitted by data sharing policies), and environment 
specification files (including \texttt{requirements.txt}) to facilitate full reproduction of the analyses and figures. 
In addition, the implemented Python library functions are available as an installable package via the Python Package Index (PyPI): 
\href{https://pypi.org/project/estimatePDF/}{https://pypi.org/project/estimatePDF/}.
\item \textbf{Author contribution:} Conceptualization: Shantanu Sarkar;
  Methodology: Shantanu Sarkar, Dexter Cahoy; 
  Software and Coding: Shantanu Sarkar;   
  Formal analysis and investigation: Shantanu Sarkar; 
  Data extraction and validation: Mousumi Sinha; 
  Supervision and Guidance: Dexter Cahoy; 
  Writing – original draft preparation: Shantanu Sarkar; 
  Writing – review and editing: Shantanu Sarkar, Mousumi Sinha, Dexter Cahoy.
\end{itemize}
%%=================================================================
%\vspace{1.5em}  % adjust value as needed
\newpage
%%=================================================================
\bibliography{References}% common bib file

\end{document}